\documentclass[10pt,letterpaper]{article}
\usepackage[top=0.85in,left=1in,footskip=0.75in]{geometry}

\usepackage{amsmath,amssymb}

\usepackage{changepage}

\usepackage[utf8x]{inputenc}

\usepackage{textcomp,marvosym}
\usepackage[numbers]{natbib}
\usepackage{graphicx}
\usepackage{nameref,hyperref}

\usepackage[table]{xcolor}

\usepackage{array}

\newcolumntype{+}{!{\vrule width 2pt}}

\newlength\savedwidth

\usepackage[aboveskip=1pt,labelfont=bf,labelsep=period,justification=raggedright,singlelinecheck=off]{caption}

\makeatletter
\renewcommand{\@biblabel}[1]{\quad#1.}
\makeatother

\begin{document}
\vspace*{0.2in}

\begin{flushleft}
{\Large
\textbf\newline{Constraining Localized Vote Tampering in the 2020 US Presidential Election} 
}
\newline
\\
Christian Johnson\textsuperscript{a}
\\
\bigskip
\textbf{a} RAND Corporation, Washington, DC, USA
Correspondence address: acjohnso@rand.org
\bigskip

\textbf{Declarations of interest:} none
\bigskip

\section*{Acknowledgments}
The author would like to thank Katharina Best for her support of this research, Kelly Klima for feedback on an earlier version of this work, as well as several anonymous reviewers who provided helpful suggestions for improving the analysis and discussion. This research did not receive any specific grant from funding agencies in the public, commercial, or not-for-profit sectors.

\section*{Disclaimer}
Christian Johnson is currently serving as an Associate Information Scientist at the RAND Corporation; however, the views, opinions, findings, conclusions, and recommendations contained herein are the author’s alone and not those of RAND or its research sponsors, clients, or grantors.

%
%

\pagebreak

{\Large
\textbf\newline{Constraining Localized Vote Tampering in the 2020 US Presidential Election} 
}

\end{flushleft}
\section*{Abstract}
Evidence of voter fraud in the United States is rare and the vote-counting system is robust against tampering, but there remains widespread distrust in the security of election infrastructure among the public. We consider statistical means of detecting anomalous election results that would be indicative of large-scale fraud, focusing on scenarios in which votes are modified in in a localized setting. The technique we develop, based on standard regression analysis, makes use of the fact that vote share is correlated with demographics. We apply our method to the results of the 2020 US presidential election as a proof-of-concept, resulting in uncertainties at the few-percent level. We are able to readily detect an artificial signal of such fraud in some cases, ruling out some scenarios of localized fraud and placing constraints on other scenarios.
\section{Introduction}
Belief that elections are free and fair is crucial to the functioning of democracy, yet public confidence in the integrity of United States elections appears to be on the decline \cite{reinhart_faith_2020}.
In some cases, this decline may stem from concerns about efforts by foreign states to infiltrate state-level election databases, as occurred in the 2016 presidential election \cite{noauthor_russian_2019}.
Others may believe in theories that allege the existence of significant levels of voter fraud (beliefs that are also promoted by foreign influence campaigns) \cite{albertson_conspiracy_2020,nakashima_us_nodate}.
Distrust in the outcome of elections has occurred in tandem with a decline in trust in other public institutions, such as the press --- part of a larger phenomenon that the RAND Corporation calls Truth Decay \cite{rich_truth_2018}.

Experts broadly agree that the American election system remains robust
\cite{ginsberg_opinion_nodate}.
A widely-cited report from the Brennan Center for Justice, for instance, concluded that voter fraud in previous US elections was virtually nonexistent, at a level of less than 0.0025\% per vote cast \cite{levitt_truth_2007}.
And because US elections are administered by thousands of independent counties, it is difficult for a single cyber campaign to compromise a significant number of votes.
Indeed, prior to the 2020 US presidential election, there was concern that so-called \textit{perception hacks} -- the spread of the false idea that voting infrastructure has been compromised, even when it remains intact -- could be a bigger problem than real cyberattacks \cite{sanger_perception_2020}.

Predictions that the voting infrastructure would remain secure were proved correct in 2020, as widespread cyberattacks and disinformation campaigns failed to materialize, and arrays of government officials declared the election infrastructure to be secure against manipulation and fraud \cite{noauthor_joint_nodate,corasaniti_times_2020}.
Despite these assurances, the outcome of the election continues to be viewed with deep skepticism by a significant portion of the US population \cite{corasaniti_times_2020}.
This skepticism is, in part, based on the fact that mail-in votes, which made up a significant part of the vote total in several key states, skewed Democratic. 
Because they tended to be counted after in-person votes, it appeared to some casual observers that the vote counting was being engineered to assure the success of the Joe Biden at the expense of Donald Trump.

One approach to shoring up public trust in the election results is to perform statistical tests on the election results to search for signs of manipulation. 
In a democracy, the results of an election, including any potential manipulation, must -- by design -- be made public.
Therefore, the presence or absence of statistically anomalous results in an election can be verified by any observer, requiring no inherent trust in any particular authority.
In the current era of widespread distrust, it is therefore interesting to consider what statistical tests are of most relevance to the 2020 US presidential election. 

There are a variety of methods described in the literature to detect anomalous vote tallies, typically under the title of "election forensics".
One method that has enjoyed a recent surge in popularity \cite{lacasa_election_2019} is to apply Benford's Law to vote counts across precincts or other subdivisions. 
Benford's Law is an empirical statement about the distribution of the leading (or next-to-leading) digit in many large, real-world datasets.
In these datasets, the leading digit is most often (about 31\% of the time) a 1, compared with the distribution expected from a random sample of numbers (about 11\%). 
Pseudorandom numbers generated by humans rarely follow such a distribution, meaning vote tallies that have been artificially generated, or have been tampered with, tend to be detectable by applying Pearson's $\chi^2$ test or similar statistical tests to the distribution of leading digits.
An analogous technique can also be applied to the next-to-leading digit, which can to be sensitive to manipulation even if the manipulator is aware of Benford's Law for the leading digit.
Benford's Law and related tests (identifying an anomalous number of 0s or 5s in the leading turnout percentage digits, for example) have been used to analyze elections in proto-authoritarian countries such as Turkey and Brazil with some success, although it should be noted that the use of these tests are not without criticism \cite{deckert_benfords_2011}.
Benford's law can be readily misunderstood or misapplied; results from the improper use of Benford's law have already been circulated as evidence for fraud in the 2020 US presidential election \cite{mebane_jr_inappropriate_2020}.

A related technique has been applied by Kobak, Shpilkin, and Pshenichnikov to a series of Russian elections and referenda \cite{kobak_integer_2016, kobak_putins_2018, kobak_suspect_2020}.
In this methodology, the turnout percentage at each polling location is computed, and anomalies are searched for in the histogram of percentages.
Russian elections consistently produce turnout histograms that show an anomalously high number of precincts with integer turnout percentages (e.g., 80\% turnout instead of 80.7\%).
The anomalous spikes can easily be explained by polling locations choosing a turnout percentage and then working backward to determine the number of votes that are needed, but are hard to explain otherwise -- in other words, these spikes are a clean signal of fraud.

Other statistical methods include analyzing the 2-dimensional distribution of turnout and vote share across all the voting precincts in an election.
This method is sensitive to ballot-stuffing, which introduces a correlation between turnout and the fraction of votes for the ruling party that can be detected.
Extreme cases of ballot stuffing can easily be identified, even by eye, using this technique \cite{kobak_statistical_2012}, because the ballot stuffing will create a bimodal distribution. 
Related methods have been researched as well, such as the use of neighboring precincts as a control to correct for demographic differences between precincts \cite{klimek_election_2017}.

Less obvious cases of fraud can be identified with a maximum-likelihood framework that fits the data with a multi-modal distribution that models fraud occurring together with free and fair voting. 
Mebane et al. combined these kinds of maximum-likelihood techniques with digit-based techniques in an R-based software package called \textbf{eforensics}, that can also perform other kinds of analysis, such as geographic clustering, on top of the results \cite{hicken_guide_2017}.
As they point out, applying multiple independent statistical tests with different methodologies and assumptions is helpful when attempting to reduce the number of false positive and negative results.
An analogous tool from Montgomery et al., Bayesian additive regression trees (BART) \cite{montgomery_olivella_potter_crisp_2017} uses a machine learning approach to assign 'fraud scores' to historical elections around the globe. 

Cottrell et al. \cite{cottrell_exploration_2018} introduced an interesting regression-based method for studying the veracity of vote fraud allegations in the aftermath of the 2016 US Presidential and Senate elections based on regression of demographic variables onto vote share and turnout results.
They regress several demographic variables, inspired by findings from political science research, onto vote share and turnout differences between the 2012 and 2016 elections. 
Fraud in the 2016 election was alleged to have occurred due to widespread non-citizen voting, so their method was aimed at uncovering the correlation between the presence of non-citizens and vote share or turnout.

Zhang et al \cite{zhang_election_2019} use a Random Forest model to detect ballot-box stuffing and vote stealing in Argentinian elections.
Their model appears to be highly accurate at detecting different kinds of fraud, given a synthetic training dataset.
However, their model relies quite heavily on turnout rates, compared to vote share -- which makes sense, given the fact that voting in Argentina is compulsory for most adults, and therefore relatively small differences in turnout are likely to be quite indicative of manipulation.
Because voter turnout is highly variable across US geographies and demographics, and because turnout incentives vary between states and elections, this same approach is not likely to be as effective for US elections.

This landscape of existing statistical tests is more than capable of demonstrating large-scale vote tampering, which often occurs in favor of the ruling party for obvious reasons.
Yet these methods methods, which generally involve studying an ensemble of results from an election, are not ideal for identifying the kinds of tampering most germane to recent United States presidential elections, where the vote counting takes place at the state and local, not federal, level and is therefore largely outside the control of the ruling party.
Instead, the biggest concerns for US elections are related to malign behavior that is geographically localized.

In particular, the nature of the Electoral College means US presidential elections can plausibly hinge on the outcome in a small number of individual counties, regardless of the national popular vote.
In other words, a somewhat small amount of fraud in the right counties could, in a close election, tilt the results in a way that is not immediately apparent from studying a distribution of data drawn from many counties.\footnote{After all, the 2016 US presidential election was won by a margin of only about 80,000 votes distributed across three states}
The most extreme case -- where the result of the presidential election hinges on the results in a single county -- is both plausible, and very likely undetectable using methods that rely on modeling large distributions.

The goal of this paper is to present a statistical method for identifying anomalous voting results that are geographically localized (as has been  alleged by some with respect to the 2020 US presidential election).
Our method, which is a straightforward application of demographic regression, is admittedly quite simple. 
Nevertheless, as we show below, it is sufficient to rule out some specific fraud scenarios that have been alleged with regards to the 2020 election that are more difficult to identify with other methods in the literature.
Our method is particularly well-suited to countering claims of fraud because of its geographic flexibility: skeptical citizens may have faith that the vote in certain areas of the country were tainted by fraud, while others were unmarred.
A voter in Texas, for instance, may believe that the Texas vote was secure (due to their first-hand experience, their faith in local and state government officials, or because of the voting laws and regulations Texas has implemented), but be skeptical about the voting process in Michigan.
Using our technique, we show that even a limited training dataset, such as the plaintiff states in the court case \textit{Texas v. Pennsylvania}, can be used to constrain the level of tampering in multiple other locations.

We caution that our statisical model (like any election forensic model) simply makes inferences -- anomalous results are not necessarily the result of voter fraud or manipulation. 
Instead, the user should look at the results in context of other data, such as the results of other statistical models, geography, or local events.
Isolated, highly anomalous results for which there is no straightforward explanation certainly warrant a closer look, but should not be taken as direct proof of malfeasance. 

The analysis we present below builds previous efforts in a few important ways.
First, we do not limit ourselves to demographic variables that we know \textit{a priori} to be correlated with voter behavior, instead using several hundred variables and using an Elastic Net to handle them.
Second, unlike Cottrell et al., we are not interested in the actual values of the parameters determined via regression.
Our modeling goal is simply to build an empirical vote model that is capable of detecting fraud, irrespective of which group may have committed it.
Finally, our method is adapted to search for localized fraud at the few-county level, which has been alleged to a greater extent in the 2020 cycle than in 2016.
Although we focus here on elections in the United States, the techniques described here could also presumably be applied to other countries with two-party systems as well.

The structure of this paper is as follows.
We first describe the Elastic Net model used to identify election anomalies and the demographic data that it is trained on.
Next we apply our technique to the 2020 US presidential election and several derived test cases, finding no conclusive evidence for anomalies consistent with fraud.
We then test the model on tampered versions of the 2020 election results in order to measure its sensitivity, finding that our model is able to correctly identify the level of fraud in some, but not all, cases.
Lastly, we detail our conclusions and recommendations.

\section{Modeling Localized Vote Tampering}\label{chap:two}
We consider the problem of localized vote tampering; that is, most of the vote across the country is free and fair, but in a small ($\mathcal{O}(1)$) number of counties, the results are modified.
This could be due, perhaps, to a cyberattack that is able to destroy voter registration data for one particular political party, or is somehow able to alter votes after they have been cast.
Perhaps a rogue election official is able to destroy boxes of ballots that were cast in her precinct, or stuff the boxes with ballots in favor of her particular candidate.
The localized scenario has indeed been alleged by some observers of the 2020 US presidential election, though it is not the only scenario alleged.
For this reason, we emphasize that the results of our method should be viewed in context with other, unrelated statistical tests, in order to get a holistic view of the election results.

Our central hypothesis for detecting this type of tampering is that some voting characteristics, particularly vote share (the fraction of votes cast for one political party), are determined in large part by demographic and geospatial indicators.
As we will show, this hypothesis does indeed appear to be true, at least from an empirical perspective.
Given this hypothesis, it is a straightforward matter to identify counties whose vote share are outliers by simply comparing the output of the model to the reported results.

Like other election forensics methods, our model is applied to the results of the election itself, and is used to check self-consistency, rather than making future predictions.
This self-consistency check can be performed in two ways: either a global fit can be performed using all the counties, after which the residuals are computed, or counties that are in question can be blinded from the model, which is fit to the remaining data. 
The blinded model is then used to predict the vote share in the questionable counties, and the residuals are computed as before.
The advantage of the blinded model is that it will be more sensitive to anomalous results, since the anomalous results will not bias the model, at the expense of having less training data available, and therefore potentially a poorer fit quality.
We employ both methods here.

If data contains extreme outliers, it is often recommended that a more detailed robustness analysis be performed to understand the sensitivity of the regression parameters to the outliers. 
Our simple model, described in more detail below, seems to work well enough for our purposes, and even in a highly-tampered election, outliers are likely to be fairly modest (see Section \ref{sec:sensitivity}); be more advanced methods for determining the robustness of the fit to outliers could be performed in future work. 

To build our model, we use an Elastic Net \cite{zou_regularization_2005}, which is a linear regression model with a penalty scheme applied to its coefficients that results in a sparse representation of the inputs.
The Elastic Net is well-suited for our situation where there are more features than samples in the data, as well as many highly-correlated features, both of which are true in our case.
Looking at the problem through the lens of machine learning, building our model is a straightforward regression task, where we seek to tune the linear regression coefficients by training on the data in such a way as to minimize the difference between the model predictions and the official results.

The model is trained to minimize the mean-squared error in predicting the Republican vote share $V$, which is defined for our purposes as the number of Republican votes divided by the total number of Republican and Democratic votes.
Note that this definition exclude third-party voters, treating them the same as non-voters.\footnote{Third-party ballot access can vary significantly from election to election, and from state to state, adding extra complexity to the model. Future work could explore the impact of third-party voting data on model performance, but as a proof-of-concept application, this initial model was kept as simple as possible.}

We use 5-fold cross-validation to select the L1-ratio\footnote{The Elastic Net applies both an $\ell_1$- and $\ell_2$- norm penalty, and the L1-ratio controls the mixing of the two: an L1-ratio of 0 has no  $\ell_2$-norm penalty, while an L1-ratio of 1 has no $\ell_1$-norm penalty.} and $\alpha$\footnote{$\alpha$ controls the overall amount of regularization applied: an $\alpha$ of 0 corresponds to ordinary linear regression.} hyperparameters automatically.
Our model is implemented in Python 3.8.2 using the {\texttt{scikit-learn}} library, version 0.23.2.

\subsection{Data Description}
The inputs to our model are demographic features, such as racial makeup, average education level, or mean income, computed at the county level, as well as vote share in the 2012 and 2016 presidential elections, computed in the same way as before.

Our demographic variables were taken from the 5-year estimates of the American Community Survey (ACS). 
In particular, we used the DP02, DP03, and DP05 datasets (selected social, economic, and demographic characteristics, respectively) \cite{us_census_bureau_american_2016}.
Duplicate features were removed (e.g., if a feature was present in both DP02 and DP03, we used only one copy of it in our training dataset), as were features that simply described the margin of error of other features, as well as features with missing data for some counties.\footnote{Missing data would have to be imputed in order to be used as a training feature, which would add additional assumptions to our model. Future work, of course, could identify more robust ways of dealing with missing data.}
We used the 2019 ACS estimates when fitting to the 2020 election (2020 estimates were not yet available at the time of the analysis). 

Our election results came from two sources: data for the 2012 and 2016 presidential elections had been acquired previously from GitHub user \texttt{tonmcg}\cite{mcgovern_united_2016}, while the 2020 results were obtained from the US Election Atlas \cite{leip_dave_2020}. 
We matched the election results in each county with the appropriate demographic data by indexing each county by its FIPS code.
Raw vote counts were also translated into vote shares. 
We excluded the state of Alaska entirely from our analysis, as Alaska does not administer its elections at the county level. 

As a result of the data-cleaning, we ended up with a total of 672 different demographic variables in total. 
They, along with the code to reproduce this analysis, are available on the Dataverse entry for this paper.\footnote{{\tt https://doi.org/10.7910/DVN/5UOPWF}}

\section{Application to US Presidential Election Results}\label{chap:three}
\begin{figure*}
\centering
\includegraphics[width = 6.5in]{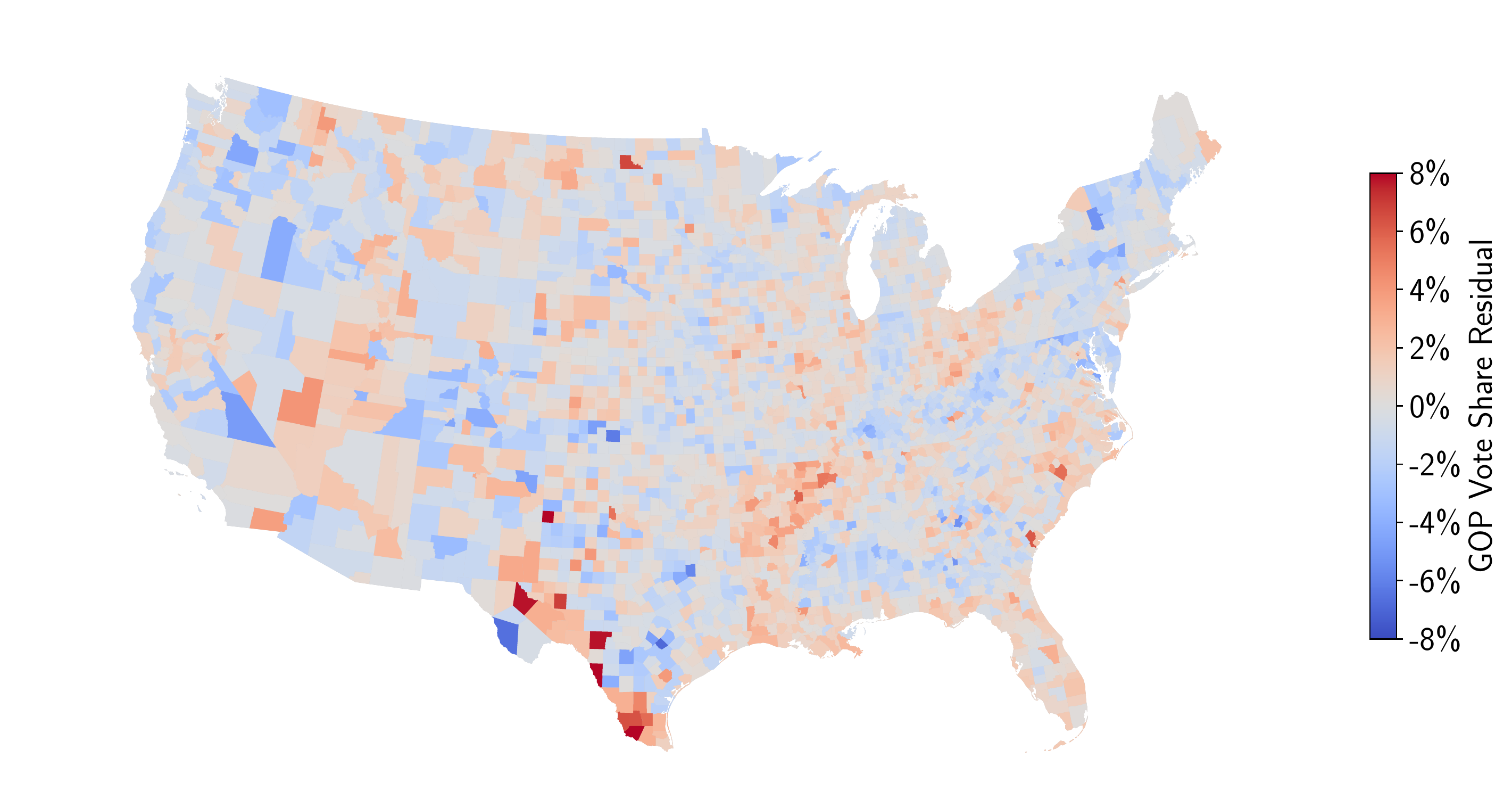}
\caption{{\bf Vote share residuals after Elastic Net fit.}
Residuals in the contiguous US after a global fit of the features to the election results in all counties. The main errors appear to be clustered in South Texas, where several counties voted significantly more Republican than expected. Because of the clustered nature of these counties, and their unique geographic circumstances, we conclude that the anomalies are likely due to some feature outside our training set.
}
\label{fig:mse_comp}
\end{figure*}

We began by performing a global fit (across all counties) to the vote share of the 2020 election results with our Elastic Net.
Cross-validation selected an L1 penalty of 0.1 and an $\alpha$ value of 0.0016. 
We find the root mean square (RMS) residual for the global fit was 1.5\% for the 2020 election.
There appear to be few geographic correlations among the residuals, as can be seen in Figure \ref{fig:mse_comp}.

Several remarks are in order about this result.
First, although our RMS residual is fairly small in absolute terms, it is relatively high compared to the known level of voter fraud.\footnote{For example, it compares quite unfavorably to the results obtained by the Brennan Center for Justice analysis cited earlier, which claimed an upper limit on fraud of 0.0025\%}
Therefore the model is unable to detect even record-setting amounts of individual voter fraud. 
This is an inevitable trade-off associated with a method that relies entirely on aggregate data.

Second, the error is similar in magnitude to the margin of victory in each of our holdout states.
Regression alone therefore cannot determine the overall winner in any of the swing states with high certainty.

However, because we are specifically concerned with the scenario of fraud that is localized in a single county, proportionally more votes would have to be altered in the single county than for the state as a whole in order to change the result in a particular state.
The question of interest here, then, is whether the scale the prediction error given above is sufficient to constrain localized fraud that capable of flipping an entire state.
As we show later, the accuracy does indeed seem to be sufficient, at least in some scenarios.
In any case, any county with a vote share different from the prediction by more than a few multiples of the RMS error certainly warrants a closer look.

Much of the remaining uncertainty that prevents our model from fitting better is presumed to come from highly granular data that was not available to us, or would require new methods to analyze.
Consider, as a thought experiment, two counties with similar demographics and prior voting patterns.
The model would make very similar vote share predictions for the outcome of the presidential election in these two counties.
In one of these counties, however, the local newspaper endorses the Democratic candidate, while the local paper endorses the Republican candidate in the other county, resulting in a difference in vote share of a few percentage points between the two counties.
This kind of highly granular data is difficult to come by, although it surely impacts the quality of the regression.
Future work might consider scraping geotagged social media data or other rich, hyperlocal datasets in order to capture these kinds of variation, and thereby reduce the remaining uncertainty of this kind of regression analysis.

We next consider the question of how, when considering the deviation of the model prediction to the data, to distinguish between fraudulent votes and a poorly-fit model.
We propose a two-pronged test: first, the results should be statistically significant, and second, there should be no indication of a missing demographic variable that might account for the anomaly.

To determine statistical significance, we first fit a normal distribution to the overall distribution of prediction errors, and determine the number of standard deviations $\sigma$ that each county is from the mean of 0. 
However, with several thousand counties to look at, and without \textit{a priori} knowledge of where fraud may have taken place, the local significance of any individual result matters less than the global significance after accounting for the look-elsewhere effect.\footnote{Here we are referring to the fact that in a large sample of normally-distributed random variables, it is expected that at least some samples will be far from the mean, simply due to pure chance. Flipping a coin five times and landing on heads every time is relatively unlikely, but if a thousand people flip a coin five times, it is not as unlikely that at least one of them will see five heads.}

We determine the global significance of each county's result with a Monte Carlo study.
First, the distribution of residuals is fit with a Gaussian distribution to find an estimate for the width of the distribution (which is fairly insensitive to outliers). 
Our choice of a Gaussian distribution is empirically motivated: the Poisson-distributed errors that might be expected from the counting statistics associated with each county's vote are somewhat smaller than the observed errors,\footnote{The distribution of the number of observed events in a finite time interval, given some average rate of events, is given by the Poisson distribution, which has a width equal to the square root of the number of expected events. 
For a county with a mean of $10^4$ votes, the intrinsic level of uncertainty due to this effect on the number of votes is therefore about 1\%.} and there does not appear to be any correlation between county size and prediction accuracy, as would be expected if the model were accurate enough to be limited by the intrinsic uncertainty of counting statistics.

With the width of the distribution in hand, we simulate a random set of results $10^6$ times by drawing 3,112 normally-distributed random numbers (the number of counties in our sample after excluding Alaska). 
The most extreme value found from each Monte Carlo run is saved, and the distribution of the resulting values is used as an estimate of the global p-value distribution for a single residual in our fit.
We convert the global p-value to a significance (in units of $\sigma$) in the standard way, by computing the inverse quantile function with the p-value as its argument.
Rejecting the null hypothesis of a non-fraudulent election then simply requires identifying the most-anomalous county and checking whether the global significance is above or below a threshold of interest.
We set our threshold at 4$\sigma$, corresponding to a probability of about 1 in 15,000 that the result occurs due to random chance.

This statistical test, of course, clearly does not determine whether the entire distribution of results is consistent with the null hypothesis.
In other words, we are looking for individual outliers by assuming that the vast majority of the counties in our sample had free and fair elections.
If many counties showed anomalous results, other existing election forensics methods could readily be employed.

The top 10 most anomalous counties are presented in Table \ref{tab:topten}.
Although the bulk of the results shown here appear to be consistent with the null hypothesis (few statistically significant anomalies), it is clear that there are several counties, particularly those close to the Texas-Mexico border, that voted far more Republican than predicted by the model.
Because of both the level of statistical significance and the geographic clustering, we posit that there was a further effect in these southern counties: either the Hispanic population near the border shifted more towards the Republican party than has been previously appreciated, or Democratic-leaning Hispanic voters were discouraged from voting for some reason related to the border.
Nevertheless, these results do not appear to indicate tampered or hacked votes, which, as noted previously, would be expected to be geographically isolated and not plausibly explained by demographic factors. 

\begin{table*}[!ht]
    \centering
    \begin{tabular}{c|c|c|c|c|c}
         \textbf{County} & \textbf{GOP Vote Share} & \textbf{Prediction} &
         \multicolumn{2}{c}{\textbf{Significance}} \\
          & & & \textbf{Local} & \textbf{Global} \\
         \hline \hline
Starr County, Texas & 47.5\% & 34.2\% & 10.4$\sigma$&-- \\
Maverick County, Texas & 45.2\% & 34.2\% & 8.7$\sigma$&-- \\
Parmer County, Texas & 81.4\% & 73.0\% & 6.6$\sigma$&-- \\
Reeves County, Texas & 61.8\% & 54.0\% & 6.1$\sigma$&-- \\
Edwards County, Texas & 84.2\% & 76.4\% & 6.1$\sigma$&-- \\
Upton County, Texas & 87.4\% & 80.5\% & 5.5$\sigma$&3.8$\sigma$ \\
Caldwell County, Texas & 54.6\% & 61.5\% & -5.4$\sigma$&3.8$\sigma$ \\
Benson County, North Dakota & 57.1\% & 50.4\% & 5.3$\sigma$&3.6$\sigma$ \\
Presidio County, Texas & 33.0\% & 39.7\% & -5.3$\sigma$&3.5$\sigma$ \\
Zapata County, Texas & 52.7\% & 46.2\% & 5.1$\sigma$&3.3$\sigma$ \\
    \end{tabular}
    \caption{Ten most highly anomalous counties among the entire US, except Alaska. We do not compute a global significance for counties with anomalies more than 6$\sigma$, as the Monte Carlo simulation would be too intensive; these can readily be assumed to be statistically significant. The results strongly suggest some unaccounted-for variable is driving voter behavior in south Texas counties, which we hypothesize in the text is the presence of the nearby border.}
    \label{tab:topten}
\end{table*}

\subsection{Case Study: \textit{Texas v. Pennsylvania}}\label{sec:blinded}

How, exactly, would the methodology presented here be used to convince an election skeptic that significant amounts of fraud did not take place in 2020?
One plausible mechanism is this: a skeptic trusts the election integrity of their own state, but is mistrustful of other states.
As a concrete example, consider the court case \textit{Texas v. Pennsylvania}, in which the Attorney General of Texas sued the states of Pennsylvania, Wisconsin, Michigan, and Georgia over (in part) alleged voting irregularities\cite{paxton_texas_2020}. 
Texas was later joined by the attorneys general of 17 other states (as well as by several members of Congress) in an \textit{amicus} brief. 
Presumably, the plaintiffs in this case would assert that the vote in their own states were without irregularity or fraud (otherwise, they would not have limited their suit to the four swing states in question).

The lawsuit makes clear that fraud is alleged to have taken place in order to tilt the states towards Joe Biden, and away from Donald Trump.
It also specifically names several high-population counties (Wayne County, MI is named many times, and Dane and Milwaukee Counties, WI are also brought up) where alleged misdeeds took place.
We are therefore presented with a clear scenario to test our methodology: training a model only on data from the plaintiff states (which are presumably free of fraud), do the defendant states demonstrate any anomalies in their voting patterns, particularly ones that show a surprisingly high rate of Democratic votes in certain high-population counties?
This experiment also addresses one potential objection that could be made to the methodology presented above, which is that any manipulation or fraud that may have occurred will decrease the quality of the fit, thereby reducing the statistical significance of any anomaly.
By blinding the model to the defendant states, we achieve a result that will be unbiased by any potential fraud.

We trained our model in the exact same way as before on the data associated with the plaintiff states only. 
In this case, the cross-validation process again found an L1 penalty of 0.1, and found an $\alpha$ of 0.0043. 
We find an average RMS prediction error of 1.6\% among the defendant states, while the RMS prediction error for the plaintiff (training) states was 1.5\%.
Promisingly, this indicates that our model does not appear to be overfitting.

\begin{figure*}
\centering
\includegraphics[width=3.5in]{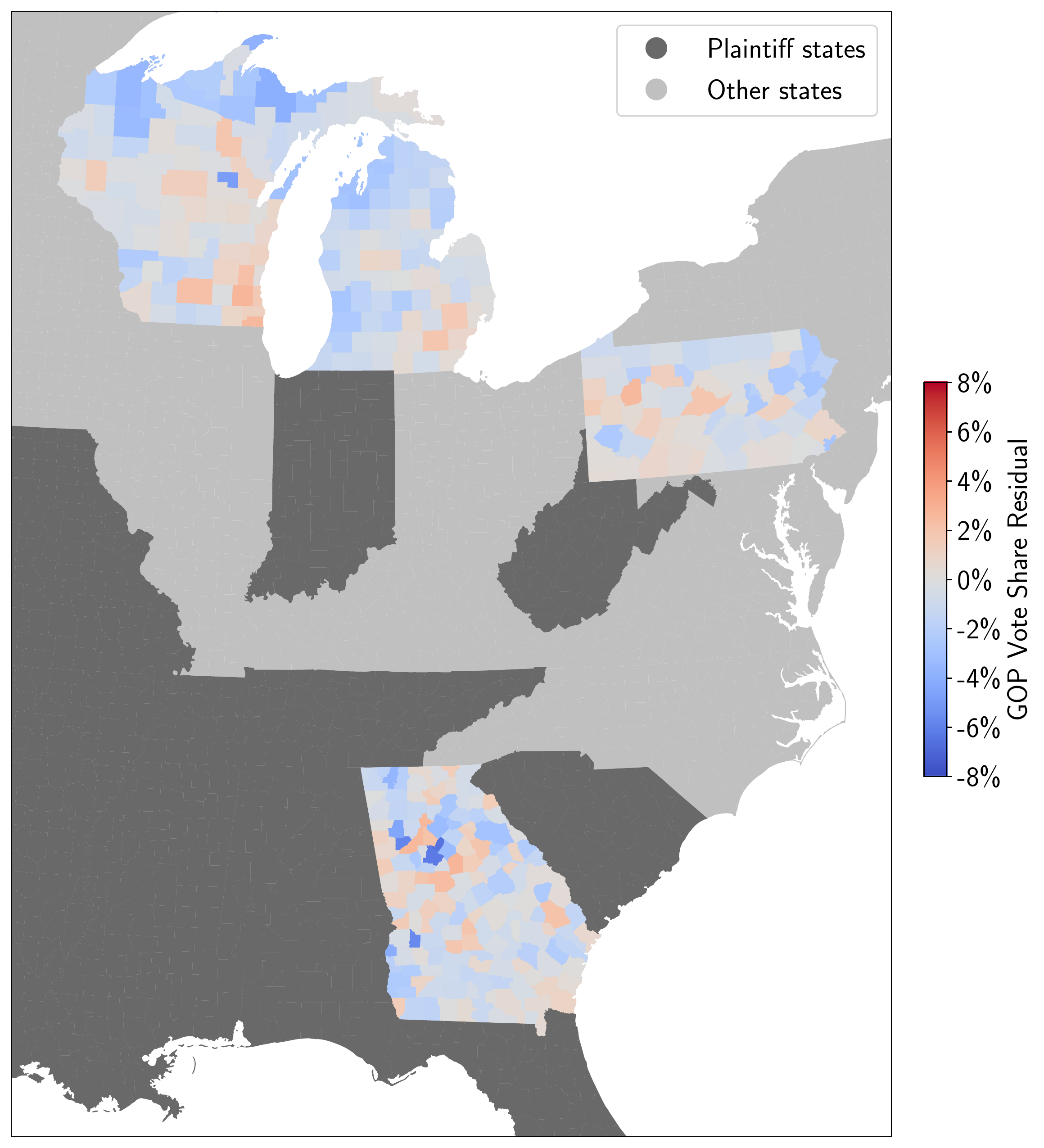}
\caption{{\bf Residuals in defendant states.}
Republican vote share residual (actual vote share minus predicted vote share) in the defendant states, shown in color, after fitting the regression model to the plaintiff states, shown in dark gray.
There is little evidence of statistically significant anomalies in the Midwest; indeed, the most contested counties, such as Wayne County, MI, appear to be well-modeled by the regression.
Several counties in Georgia have residuals of moderate statistical significance, though this appears to be unrelated to fraud, as described in the text.
}
\label{fig:defendant_states}
\end{figure*}

The results of this experiment are depicted in the choropleth map in Figure \ref{fig:defendant_states}.
There are few anomalies of great significance in the defendant states, as can be seen in Table \ref{tab:defendant_states}.
A handful of counties in Georgia were more than a few standard deviations away from the model prediction.
Again, however, there appears to be geographic clustering to this effect, which indicates that some demographic or civic factor was not taken into account. 
Rockdale and Henry Counties (the most anomalous) indeed were noted in local media for a rapidly changing electorate\cite{hallerman_georgia_2020}.
Because the model only has a snapshot of demographic data, this could indicate that incorporating time-series features would improve its performance.

\begin{table*}[t]
    \centering
    \begin{tabular}{c|c|c|c|c|c}
         \textbf{County} & \textbf{GOP Vote Share} & \textbf{Prediction} &
         \multicolumn{2}{c}{\textbf{Significance}} \\
          & & & \textbf{Local} & \textbf{Global} \\
         \hline \hline
Rockdale County, Georgia & 29.4\% & 36.0\% & -5.3$\sigma$&4.1$\sigma$ \\
Henry County, Georgia & 39.7\% & 45.9\% & -5.0$\sigma$&3.7$\sigma$ \\
Douglas County, Georgia & 37.3\% & 43.1\% & -4.7$\sigma$&3.3$\sigma$ \\
Webster County, Georgia & 53.9\% & 59.5\% & -4.5$\sigma$&3.1$\sigma$ \\
Menominee County, Wisconsin & 17.6\% & 22.4\% & -3.9$\sigma$&2.1$\sigma$ \\
Paulding County, Georgia & 64.7\% & 69.2\% & -3.6$\sigma$&1.6$\sigma$ \\
Quitman County, Georgia & 54.9\% & 59.0\% & -3.3$\sigma$&1.0$\sigma$ \\
Marquette County, Michigan & 44.3\% & 48.4\% & -3.3$\sigma$&0.9$\sigma$ \\
Keweenaw County, Michigan & 56.2\% & 60.0\% & -3.1$\sigma$&0.6$\sigma$ \\
Whitfield County, Georgia & 70.6\% & 74.2\% & -2.9$\sigma$&0.3$\sigma$ \\

    \end{tabular}
    \caption{Ten most highly anomalous counties among the defendant states in \textit{Texas v. Pennsylvania}, after training the ElasticNet model on the plaintiff state data. Several counties in Georgia voted significantly more Democratic than predicted by the model. As noted in the text, there are a variety of explanations for this fact that do not rely on fraud, which we hypothesize would likely manifest differently.}
    \label{tab:defendant_states}
\end{table*}

Interestingly, inspection of Figure \ref{fig:defendant_states} shows that the counties under the highest scrutiny by the plaintiffs (Wayne County, MI, and Dane and Milwaukee Counties, WI) actually were slightly more Republican-leaning than the model trained on the plaintiff states predicted.
We also note that if the model predictions were perfectly accurate,\footnote{By 'perfectly accurate', we mean that the vote share is assumed to be correct in each county, while the total number of votes cast combined for both Republicans and Democrats is unchanged} Michigan and Wisconsin would still have ended up voting for Joe Biden, while Georgia and Pennsylvania would have instead narrowly gone for Donald Trump - still leading to a Biden win in the Electoral College.
Our results are therefore in tension with the claim in the lawsuit that the alleged fraud in these states was `outcome-determinative'. 

As it so happens, this lawsuit was quickly dismissed by the Supreme Court on unrelated grounds.
In a more contentious court case, however, the methodology presented here could help shore up confidence with the plaintiffs that no outcome-determinative fraud took place - unless, of course, the plaintiffs are willing to admit that similar fraud took place in their own states as well, which would bias the model into ignoring vote tampering.

\subsection{Sensitivity testing}\label{sec:sensitivity}
Although the results presented above are broadly consistent with the null hypothesis (i.e., no clear evidence of fraud), we sought to understand what such a signal would look like, and to determine what minimum level of fraud our model would be sensitive.
For the first goal, injecting a signal of fraud and then reconstructing it would give confidence that the method, despite its simplicity, is at least moderately sensitive to the type of fraud we are interested in. 
The second goal can be accomplished by injecting varying levels of signal into the data, and observing the point at which the signal becomes detectable above our 4$\sigma$ threshold.

To inject a signal of fraud, we modified the results in Wayne County, MI, by adding 70,000 Democratic votes and removing 70,000 Republican votes. 
This might be the kind of fraud expected if, for instance, maliciously-programmed voting machines had been programmed to switch votes after they had been cast.
The net 140,000 votes would not be enough to flip the results in the entire state of Michigan, but come close.
We chose Wayne County, MI for this test not only because it is named numerous times in \textit{Texas v. Pennsylvania}, but because it is the most populous county in Michigan, so hiding fraud should be easiest there -- 140,000 flipped votes in a smaller county would yield an even greater change in the vote share.
Our experiment is therefore somewhat conservative: if a malicious actor wanted to flip the state of Michigan, this scenario should be as difficult to detect as possible under our assumptions.

\begin{figure*}
\centering
\includegraphics[width = 3.5in]{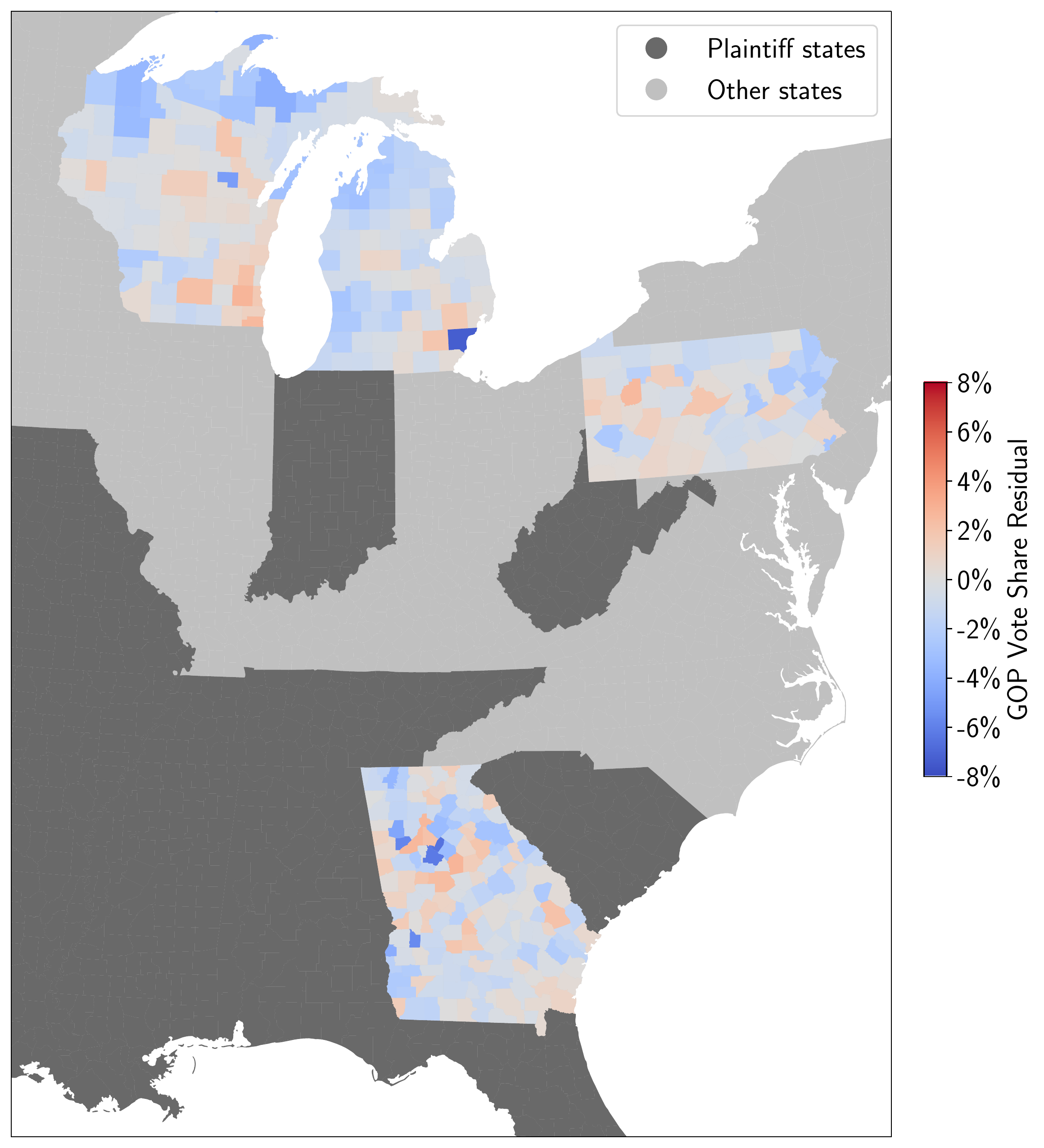}
\caption{{\bf Detection of injected fraud in Wayne County, MI.}
Residuals of the fit after introduction of an injected signal in Wayne County, MI. 
The votes in Wayne County are modified by removing 70,000 Republican votes and adding an extra 70,000 Democratic votes, which would be close to being outcome-determinative for the state of Michigan.
The isolated nature of the anomaly, coupled with its relatively high statistical significance, give us confidence that such tampering would be detected if it were enough to flip the result of a state like Michigan.
}
\label{fig:michigan}
\end{figure*}

In Figure \ref{fig:michigan}, the altered votes are immediately apparent by eye.
We find a residual of -7.3\% in vote share, corresponding to a local significance of -5.9$\sigma$ and a global significance of -4.8$\sigma$.
It is notable that our reconstructed signal compares quite favorably to the -8.1\% change in vote share that was artificially injected.
This level of significance would immediately call attention to Wayne County in post-election studies, as similarly large anomalies drew our attention to the southern-Texas counties.
We therefore rule out with some confidence this particular fraud scenario.

Next, we perform the same experiment for all counties in Michigan, Wisconsin, Pennsylvania, and Georgia with sufficient votes to flip the outcome in each state.
Specifically, we ask how much vote-flipping could have taken place in the 2020 election before anomalous results would be apparent with our model.
We consider both Republican-to-Democrat as well as Democrat-to-Republican vote flips.

The results of the sensitivity testing are mixed. 
We find that whether vote-flipping is detectable is highly dependent on three main factors: the margin of victory in each state, and the populations and vote shares of the largest counties in each state.
In a state like Georgia, with a margin of just under 12,000 votes, there are many counties where switching 6,000 votes for one party to another results in very little change in the vote share, and would therefore be undetectable.
In Michigan, with a margin of over 150,000 votes, the story is very different: there are no counties in which enough votes could have been flipped to make up the margin of victory.

\begin{figure}[!h]
\centering
\includegraphics[width=4in]{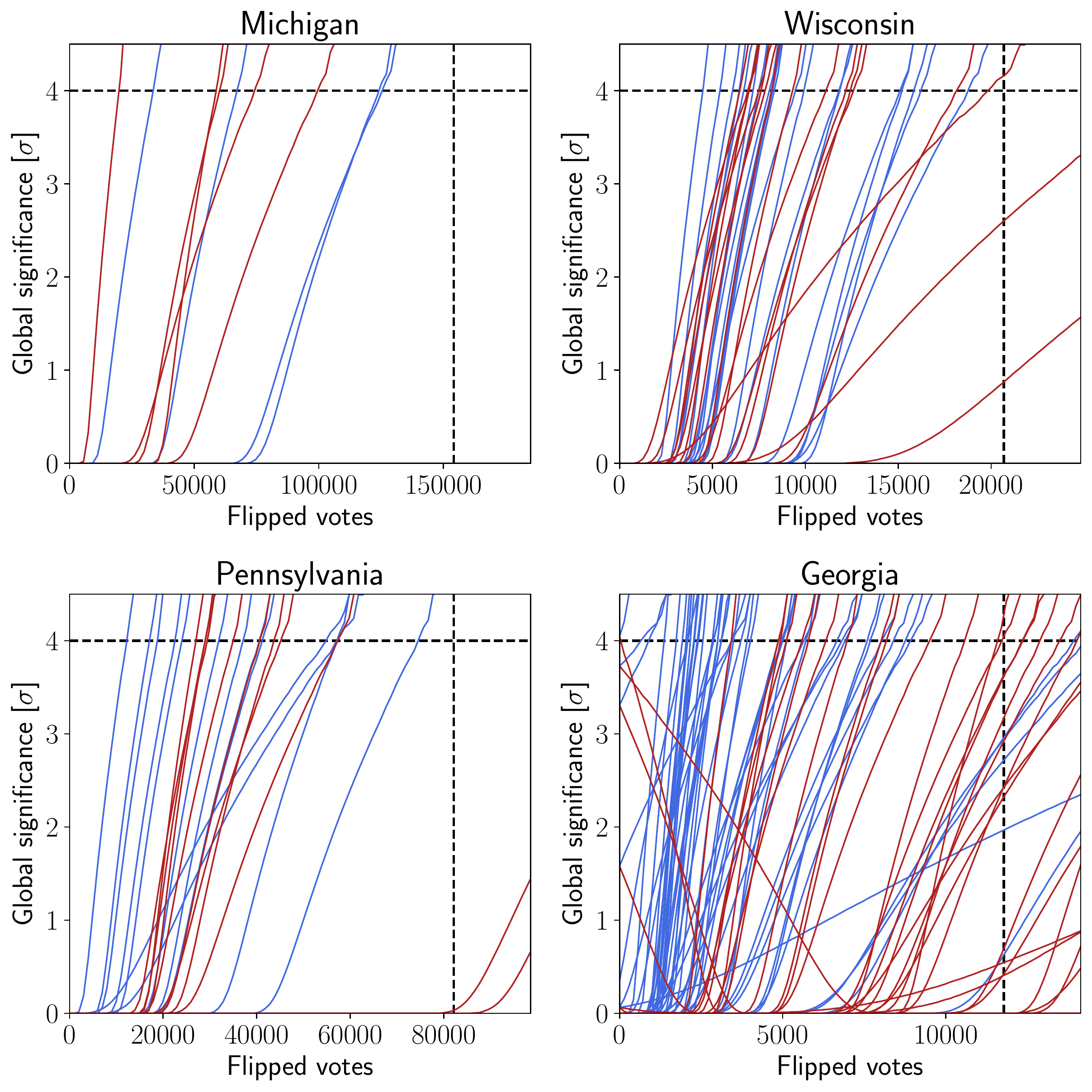}
\caption{{\bf Sensitivity testing in four swing states.}
Global significance as a function of the number of flipped votes in the four swing states considered. 
Each line corresponds to a single county; blue lines indicate flipping Republican votes to Democratic, and red lines correspond to flipping Democratic votes to Republican (using the 2020 results as a baseline). For reference, the 2020 margin of victory is shown as a vertical dashed line in each plot, and our threshold of 4$\sigma$ is shown as a horizontal dashed line.
Lines that cross below the intersection of these dashed lines correspond to counties where an outcome-determinative amount of vote-flipping could (in theory) have occurred without becoming statistically significant.
Only counties with more Democratic or Republican votes than the margin of victory are shown.
As expected, in states with larger margins of victory (i.e., Michigan), the amount of vote-flipping needed to change the statewide result would be easily detectable, as was seen in the previous section.
For a state like Georgia, with a narrow margin of victory, our method is not sensitive enough to rule out a relatively large number (19) of counties. 
}
\label{fig:spaghetti}
\end{figure}

Pennsylvania and Wisconsin lie between these two extremes, with two counties each (Philadelphia and Allegheny in Pennsylvania; Dane and Milwaukee in Wisconsin) where the statewide margin of victory is smaller than the amount of vote-flipping that would have needed to take place in order for the signal to be detectable at the 4$\sigma$ level.
The results from all four states are displayed in Figure \ref{fig:spaghetti}.

\section{Discussion}\label{chap:four}
Our results are consistent with election results that are unmarred by any kind of localized manipulation or fraud. 
However, we caution the method described here is only capable of identifying fraud at levels of a few percent -- higher than the margin of victory in several states.
This is somewhat mitigated by the fact that the population of each state is distributed among many counties, so changing the outcome of the statewide result requires a larger level of vote tampering at the county level than the state level.

Our experiment with altering votes in Wayne County, MI confirms that even the smallest amount of fraud possible to flip the state would have been quickly detected.
Further sensitivity testing revealed that in swing states with narrower margins of victory, the model can only rule out an outcome-determinative amount of fraud in most, but not all, counties.
This is not surprising; in counties with many hundreds of thousands of votes cast and a statewide margin of only a few tens of thousands of votes, the fraction of votes that would need to be flipped is only a few percent - and few-percent variation is quite expected by our regression model.

The value of this modeling method, in our view, is that it can constrain both the amount and location of potential fraud, rather than eliminating the possibility of fraud altogether.
With the simple model presented here, we were already able to strongly constrain the outcome in Michigan, and severely constrain the potential locations for fraud in Wisconsin and Pennsylvania.
More powerful models in the future, of course, may place even stronger constraints that would eliminate any likelihood of any outcome-determinative fraud.

We recognize that statistical models trained on a broad swath of data are unlikely, by themselves, to convincingly show that an election was free and fair. 
As we noted above, analyses using Benford's law have already been inappropriately applied to the election results, and it is certainly possible that regression models could also be misused to push a stolen election narrative.
But because it uses an entirely different methodology than Benford's law or similar tests, the model presented here is likely to produce results that are somewhat orthogonal to those from other kinds of statistical tests; that is, in a free and fair election, different statistical tests will give uncorrelated results, but in a fraudulent election, multiple tests should point to the same conclusion.
A large ensemble of qualitatively different statistical models, then, is likely to be not only the most sensitive to fraud, but also the most convincing.

\bibliography{references}

\end{document}